\documentclass[9pt]{article}

\usepackage[T1]{fontenc}
\usepackage[english]{babel}

\usepackage[square,numbers]{natbib}
\usepackage{color,geometry,graphicx,gensymb,amsmath,bm,wrapfig,overpic}
\usepackage[font=footnotesize,skip=5pt]{caption}
\usepackage{enumitem,calc}
\usepackage[small]{titlesec}
\usepackage{hanging}
\usepackage{upgreek}
\usepackage{booktabs}
\usepackage{multirow}
\usepackage{bm}

\definecolor{darkpastelgreen}{rgb}{0.01, 0.75, 0.24}

\title{Nanocrystalline diamond-glass platform for the development of three-dimensional micro- and nanodevices}
\author{Stoffel D.\ Janssens\footnote{Corresponding author.}~, David V\'azquez-Cort\'es, Alessandro Giussani, James A.\ Kwiecinski, Eliot Fried}
\date{}
\begin{document}
\maketitle
\vspace{-1cm}
\begin{center}
Mathematics, Mechanics, and Materials Unit (MMMU), Okinawa Institute of Science and Technology Graduate University (OIST), 1919-1 Tancha, Onna-son, Kunigami-gun, Okinawa, Japan 904-0495
\end{center}
\vskip15pt
Low-cost and robust platforms are key for the development of next-generation 3D micro- and nanodevices. To fabricate such platforms, nanocrystalline diamond (NCD) is a highly appealing material due to its biocompatibility, robustness, and mechanical, electrical, electrochemical, and optical properties, while glass substrates with through vias are ideal interposers for 3D integration due to the excellent properties of glass. However, developing devices that are comprised of NCD films and through glass vias (TGVs) has rarely been attempted due to a lack of effective process strategies. In this work, a low-cost process --- free of photolithography and transfer-printing --- for fabricating arrays of TGVs that are sealed with suspended portions of an ultra-thin NCD film on one side is presented. These highly transparent structures may serve as a platform for the development of microwells for single-cell culture and analysis, 3D integrated devices such as microelectrodes, and quantum technologies. The process is demonstrated by fabricating TGVs that are sealed with an NCD film of thickness 175~nm and diameter 60~$\upmu$m. The technology described can be extended by replacing NCD with silicon nitride or silicon carbide, allowing for the development of complex heterogenous structures on the small scale.
\vskip5pt

\noindent
\textit{Keywords:} nanocrystalline diamond, through glass vias, glass etching, laser ablation, microfabrication 
{\let\thefootnote\relax\footnote{Email address: stoffel.janssens@oist.jp (S.\ D.\ Janssens).}}
\setcounter{footnote}{0}
\vskip10pt
\begin{center}
\line(1,0){450}
\end{center}
\vskip20pt

\section{Introduction}

Nanocrystalline diamond (NCD) is a highly appealing material for a variety of applications due to its biocompatibility, robustness, and mechanical, electrical, electrochemical, and optical properties. At odds with the general misconception that diamond is expensive, diamond films can be grown at low cost with microwave plasma assisted chemical vapor deposition (MWPACVD) \cite{Achard2018}. For NCD, this can be done on large area substrates using an inexpensive precursor mixture, consisting usually of methane gas diluted in molecular hydrogen, after the deposition of diamond seeds of diameter below 10~nm \cite{Williams2007}. During growth, a boron-containing precursor can also be added to make \textit{p}-type diamond that exhibits metallic properties when heavily doped \cite{Gajewski2009,Janssens2011}. Boron-doped NCD can have a relatively wide potential window \cite{Hupert2003} which, in combination with the chemical inertness and biocompatibility of diamond \cite{Tang1995}, makes it an attractive material for electrodes \cite{Hartl2004,Alcaide2016}. During growth, dopants such as nitrogen and silicon can also be incorporated in diamond to form color centers for quantum technologies \cite{Tzeng2015,Stehlik2017}. After growth, NCD films can be processed to form two or three-dimensional suspended structures \cite{Tian2017,Abdou2018} and the surface of an NCD film can be functionalized with a variety of (bio)molecules for use in biosensors \cite{Caterino2015} and solar cells \cite{Raymakers2018}. Due to these excellent properties, NCD based devices such as micromechanical resonators of high $Q$-factor \cite{Wang2004,Sartori2018}, pressure sensors for harsh environments \cite{Adamschik2001,Janssens2014a}, tunable optical lenses \cite{Kriele2009}, biosensors that, for example, can detect influenza \cite{Matsubara2016,Siuzdak2019}, optically transparent electrodes \cite{Lim2010,Ashcheulov2017}, CO$_2$ reducing electrodes \cite{Roy2016}, superconducting quantum interference devices \cite{Mandal2011}, and conducting atomic force microscope tips are being developed \cite{Celano2018}.

For the fabrication of 3D structures in microdevices, interposing glass layers with through holes that serve as conduits for fluids or electrical connections to thin films are indispensable \cite{Esashi2008,Iliescu2012,Sugioka2014}. In recent years, these interposing layers have been made available by several companies in the form of thin glass substrates with through glass vias (TGVs). Being inexpensive, transparent, electrically insulating, chemically inert, biocompatible, of high mechanical stiffness, and reusable, glass is a natural choice for the fabrication of interposing layers \cite{Sukumaran2012}. For example, it was recently shown that the use of TGVs can lead to low-loss and high-linearity radio frequency interposers \cite{Shah2018}. Moreover, the properties of glass are strongly tunable. It can, for example, be made with coefficients of thermal expansion similar to those of semiconductor materials such as silicon and can thus be used to fabricate microdevices with minimal residual stress \cite{Kazutaka2017}. Work disclosed by Corning shows that a viable process for fabricating TGVs is based on wet etching and either laser ablation or the modification of glass with laser light \cite{Burkett2015,Jaramillo2018}, while AGC relies on a process that is based on a focused electrical discharging method \cite{Takahashi2013}. Recent work by Sato et al.\ \cite{Sato2019} shows that laser ablation is promising for the fabrication of TGVs in polymer-laminated thin glass substrates. The polymer acts as a support for the thin glass plate \cite{Sukumaran2012}.

Towards the development of devices that rely on the exceptional properties of glass and NCD, we present a low-cost process, free of photolithography and transfer printing, for fabricating arrays of TGVs that are sealed on one side with suspended portions of an ultra-thin NCD film. The resulting platform can be useful for single-cell culture and analysis when the NCD film is made porous, which can be achieved through annealing \cite{Kriele2011}, and thus, be used for nutrient or drug delivery. From work on the fabrication of robust membranes by Salminen et al.\ \cite{Salminen2019}, it is also clear that our platform has a future in the field of modeling vascular systems. When made with boron-doped NCD films, our platform can be used to construct electrodes for microfluidic channels. Alternatively, if comprised of NCD with properly arranged lattice defects, our platform can also be used for quantum technologies. It is also noteworthy that NCD can be replaced by other materials that are resistant to hydrofluoric acid (HF), such as silicon carbide or silicon nitride, and that their thermal properties, along with those of the glass that we use, allow for the fabrication of MEMS that can operate in air up to temperatures of about 400~$\degree$C.

\section{Materials and methods}
\label{section_materials_and_methods}
\subsection{Process overview}
\begin{figure*}[t]
    \centering
    \includegraphics[scale = 0.95]{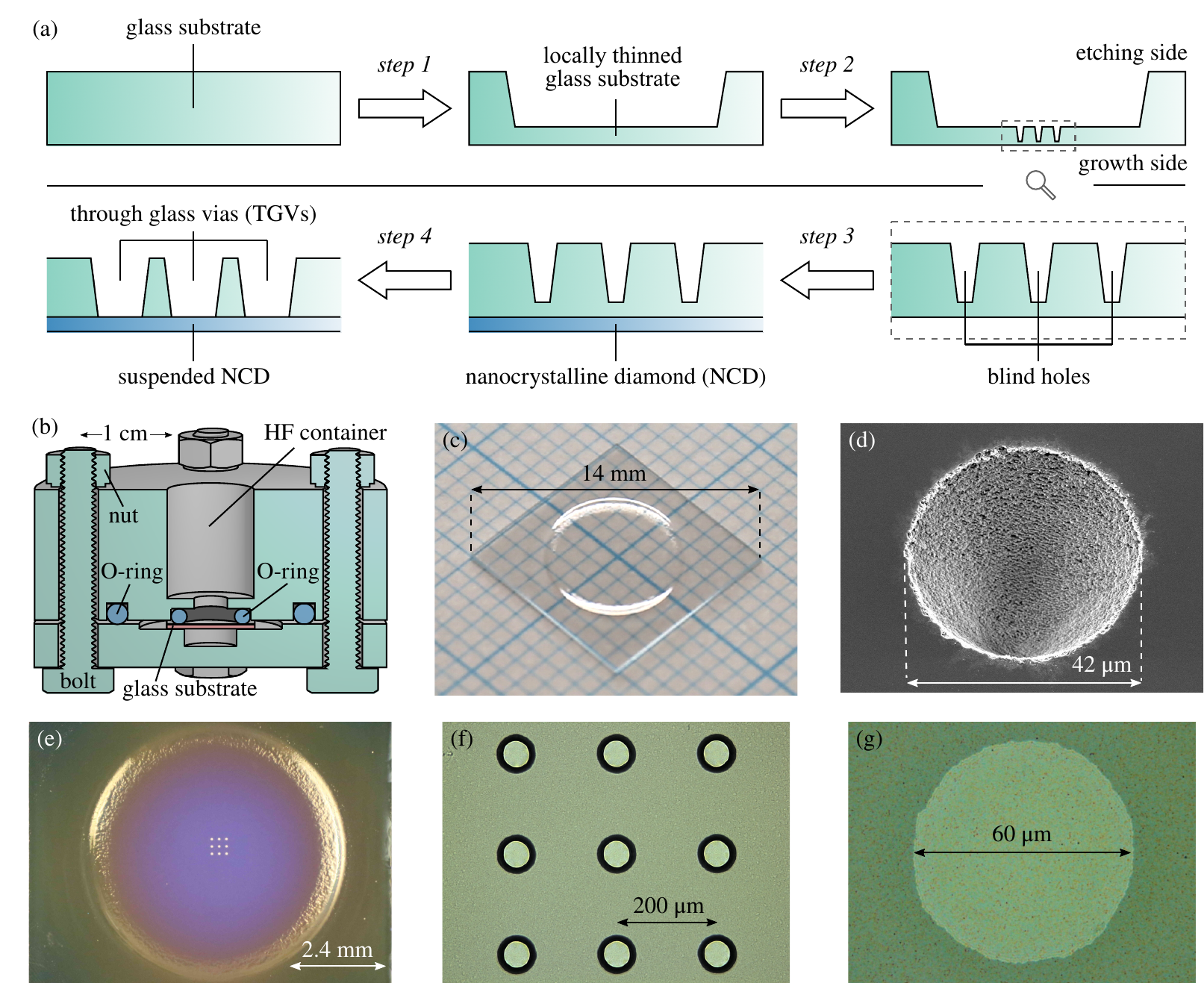}
    \caption{(a) Schematics of the cross-section of a substrate before and after steps in the fabrication process of through glass vias (TGVs) that are sealed with portions of a nanocrystalline diamond (NCD) film. In step~1, a glass substrate of approximately 200~$\upmu$m thick is locally etched by hydrofluoric acid (HF) on the etching side to a thickness of approximately 50~$\upmu$m. The non-etched portion of the substrate serves as a supporting frame for the etched portion. In step~2, blind holes of approximate diameter 42~$\upmu$m and approximate depth 40~$\upmu$m are made by laser ablation. In step~3, an NCD film of thickness less than 180~nm is grown on the growth side. In step~4, HF etching is done locally to form the NCD sealed TGVs. In the neighborhood of the TGVs, the substrate is of approximate thickness 25~$\upmu$m. (b) Schematic of the cross-section of the home-built chemical reactor used to etch glass substrates. A detailed description of the reactor and its use is provided as supplementary material. (c) Image of a $10 \times 10 \times 0.2$~mm$^3$ Lotus NXT glass substrate, laying on graph paper, taken immediately after step~1. (d) Scanning electron microscope (SEM) image of a blind hole that is made by laser ablation. The image is taken under a tilt of 25$\degree$. (e) Image of a sample, taken with the camera on the growth side, immediately after diamond growth. (f) Image of an array of NCD sealed TGVs taken with a reflecting light microscope with the objective on the etching side. (g) Image of a sample,  taken with the same reflecting light microscope but with the objective on the growth side. The circular structure represents the center TGV of Fig.~\ref{process}.f and is visible due to the difference between the refractive index of air and glass.}
    \label{process}
\end{figure*}

Fig.~\ref{process}.a shows a schematic of the cross-section of a glass substrate before and after each fabrication step. See the figure caption for a brief description of each step. A schematic of the home-built chemical reactor used to etch glass substrates appears in Fig.~\ref{process}.b and sample images at each stage of the fabrication process appear in Figs.~\ref{process}.c--g.  

\subsection{Glass substrates and etching}
\begin{table*}
\footnotesize
  \caption{Types of glass upon which nanocrystalline diamond (NCD) can be grown and their properties. All data was obtained from the suppliers.}
  \label{types_of_glass}
  \begin{tabular*}{\textwidth}{@{\extracolsep{\fill}}llllll} 
  \toprule
  \multirow{2}{*}{\textbf{Properties}} & \multicolumn{5}{c}{\textbf{Types of glass}}\\
     & Schott AF45 & AGC AN100 & Corning Lotus NXT & Corning Eagle 2000 & AGC Fused Silica\\
    \cmidrule{2-6}
    Annealing point$^{\dag}$ [$\degree$C] & 663 & 720 & 806 & 722 & 1120\\
    $\alpha^{\ddag}$ [$10^{-6}$ \degree C$^{-1}$] & 4.5 & 3.8 & 3.5 & 3.2 & 0.6\\
    Density (25~$\degree$C) [g~cm$^{-3}$] & 2.72 & 2.51 & 2.59 & 2.37 & 2.20\\
    Resistivity (250~$\degree$C) [$\Omega$ cm] & $6 \times 10^{\text{13}}$ & $4 \times 10^{\text{13}}$ & $20 \times 10^{\text{13}}$ & $0.3 \times 10^{\text{13}}$ & $0.2 \times 10^{\text{13}}$ \\
    Young's modulus  (25~$\degree$C) [GPa] & 66 & 77 & 83 & 71 & 74\\
    \bottomrule
  \end{tabular*}
    $^{\dag}$The temperature at which the viscosity of the glass is 10$^{13}$ Pa s.\\
    $^{\ddag}$Linear coefficient of thermal expansion averaged from 0 to 300~$\degree$C. For diamond, the value for this property is approximately $2 \times 10^{-6}~$\degree C$^{-1}$ \cite{Moulle1997}.
\end{table*}
To yield $10 \times 10 \times 0.2$~mm$^3$ substrates, a 200~$\upmu$m thick plate of Corning Lotus NXT glass, which is an alkaline earth boro-aluminosilicate, was diced with a Disco DAD322. The plate was purchased from the company Kuramoto. After dicing the plate, the substrates were cleaned for 20~min in acetone with an ultrasonicator. The remaining acetone was rinsed from the substrates with deionized water. The properties of Lotus NXT glass are listed in Table~\ref{types_of_glass}.

Isotropic etching of silicon dioxide-based glass is typically done with HF. The overall chemical reaction for etching silicon dioxide glass with HF is 
\begin{equation}
\textrm{SiO}_2 + 6\textrm{HF} \rightarrow \textrm{H}_2\textrm{SiF}_6 + 2\textrm{H}_2\textrm{O}.
\label{glass_etch}
\end{equation}
This, however, represents a simplification of what occurs during the etching steps of our process. Moreover, no clear mechanism for glass etching is provided in literature \cite{Spierings1993}. The etching rate of silicon dioxide-based glass with HF typically increases with temperature, hydrogen fluoride concentration, and the concentration of species other than silicon dioxide \cite{Spierings1993,Liang1987,Tay2006}. To synthesize a silicon dioxide-based glass with specific properties, well-defined amounts of oxides such as Al$_2$O$_3$, As$_2$O$_3$, B$_2$O$_3$, CaO, K$_2$O, MnO, Na$_2$O, and P$_2$O$_5$ are mixed with SiO$_2$ \cite{Spierings1993}. According to Tay et al.\ \cite{Tay2006}, these oxides react with HF to form fluorides that are insoluble in HF and which sediment onto portions of the etching surface leading to roughening. Iliescu et al.~\cite{Iliescu2005} found that adding hydrochloric acid to HF provides an effective way to dissolve these fluorides. In addition, Ceyssens and Puers~\cite{Ceyssens2009} noted that the sedimentation of these fluorides can be avoided by vertically orienting the glass surface that is being etched. They reasoned that gravity drags the fluorides away from the surface, preventing masking. 

In this work, glass etching was done in a class 1000 cleanroom at a temperature of 23~$\degree$C and a humidity of 60\%. In step~1, 0.6~ml of 48~m\% HF was used. Since the etching rate for 48~m\% HF was found to be too high for controlled etching in step~4, 11~m\% HF was used in that step.
\subsection{Laser ablation}
Blind holes were made with a LightFab system designed for selective laser-induced etching of fused silica \cite{Marcinkevicius2001,Hermans2014}. The LightFab system was  equipped with a 4~W laser. The laser light, of wavelength 1030~nm, was focused with a microscope objective to a spot of approximate diameter 1~$\upmu$m. The objective had a numerical aperture of 0.4, magnified 20 times, and had a working distance of 10~mm. The galvo scanner head of the system allowed the focal plane $F$ to move in three dimensions and with a stepper motor. The stage of the system could also move in three dimensions. The front side laser ablation technique was used to create blind holes where the substrate was etched maximally to approximately 50~$\upmu$m. Due to software restrictions, the depth $f$ of $F$, measured relative to the surface located at the etching side of the substrate, was changed with the stepper motor rather than with the galvo. We found that crack formation during ablation was minimized by using a laser pulse frequency of 500~kHz, a pulse duration of approximately 270~fs, and a write speed of 150~mm~s$^{-1}$. With these parameters, the ablation threshold of the glass surface was found to occur at $P=0.33$, where $P$ denotes the laser power scaled by the maximum laser power. The pattern that was written during laser ablation consists of 20 concentric circles each of which is separated from its immediate neighbors by 1~$\upmu$m. The largest circle was of diameter 40~$\upmu$m, whilst the smallest circle is of diameter 2~$\upmu$m, and the sequence of writing was from the circle of largest diameter to the circle of smallest diameter. To create holes, we sequentially wrote the pattern and lowered $F$ to a prescribed $f$.
\subsection{NCD growth}
During the chemical vapor deposition of an NCD film, temperature $T_s$ of the substrate during growth typically ranges from 500~$\degree$C to 900~$\degree$C \cite{Koizumi2008}. For our purposes, glasses with annealing points above 500~$\degree$C are therefore preferable. To avoid stress due to thermal mismatch between the glass and the NCD, the glass should also be designed so that its coefficient of thermal expansion is equal to that of NCD for the complete range of temperatures achieved during the growth process, a criterion that can be difficult to meet. However, such glass is not currently available commercially. Still, several types of silicon dioxide-based glass on which NCD can be grown have annealing points above 500~$\degree$C. Glasses that can be used for NCD growth are listed in Table~\ref{types_of_glass}, together with some of their salient properties. Except for fused silica, the glasses listed in this table are designed to be used in conjunction with silicon.

Preceding NCD growth, the substrate was cleaned using the first, third, and fourth steps of the RCA method. After cleaning, the surface on the growth side was seeded \cite{Williams2007} with detonation nanodiamonds of diameter below 10~nm. Although nanodiamonds are chemically bonded to a matrix of sp$^2$ hybridised carbon after synthesis, they can be separated by bead milling \cite{Kruger2005}. With a powder of such separated nanodiamonds, purchased from NanoCarbon Research Institute Co., Ltd., we next made a stable colloidal suspension using a method similar to that reported by Ozawa and coworkers \cite{Ozawa2007}. We did this by first mixing 0.1~g of the powder in 0.2~l water ($\approx 0.05$~m\%) and then ultrasonicating the mixture using an ultrasonic probe with a tip of diameter 3.2~mm and length 4.5~cm. The probe was connected to a transducer of power 100~W and frequency 20~kHz that was set to cycle on and off every second for 90 minutes. The obtained suspension was turbid but cleared after a week with the settling of particles. After mounting the substrate on a spin coater, 40~$\upmu$l of the suspension was drop cast on the substrate. Alternatives to drop casting involve submerging the substrate in the suspension or squirting the suspension on the substrate \cite{Williams2007,Tsigkourakos2012}. One minute after drop casting, the surface of the substrate was flushed for 10~s with deionized water while the substrate was spinning at 4000~rpm. This strategy was used to avoid any aggregation of nanodiamonds. Simply dipping the substrate in an abundance of deionized water also leads to satisfying results; this, however, can lead to unintentional seeding of substrate surfaces that should remain unseeded. After flushing, the substrate was dried by spinning for an additional 15~s without changing the spin frequency. 

The seeded substrate was placed in the reactor of an SDS6500X MWPACVD system on a molybdenum substrate holder of 58~mm diameter and 5.5~mm thickness. Subsequently, the gasses in the reactor were evacuated by a Kashiyama SDE90X dry pump to base pressure~8.5 Pa. Hydrogen gas and methane gas were next introduced into the reactor at respective flow rates of 294~sccm and 6~sccm. After reaching a stable operation pressure $P = 2$~kPa, the gasses were ignited into a plasma with 1.5~kW of 2.45~GHz microwaves. At these conditions, $T_s$ remained relatively far below the annealing point of Corning Lotus NXT glass, which is 722~$\degree$C, and the growth rate $r$ was on the order of 1~nm~min$^{-1}$. This value of $r$ typically occurs at relatively low values of $T_s$ and $P$. The thicknesses of the NCD films were measured during growth with a home-built interferometer. During growth, a light gray film, consisting likely of hydrogenated carbon, was unintentionally deposited on the etching side of the substrate. Within a minute, this film was removed in the reactor of a Yamato PR200 system.

\subsection{Characterization}

All photographic images were taken with a Canon EOS 5D camera and a Canon MP-E 65~mm lens. Surface profile features with lengths greater than 1~mm were taken with a DektakXT stylus profilometer and surface profile features with lengths less than 1~mm were taken with a Keyence VK-X150 laser microscope. From surface profiles that were taken with this microscope, the arithmetic average surface roughness $R_a$ was deduced. The VK-X150 was also used to measure the thicknesses of the glass substrates, taking into account the refractive index, 1.53, of the glass substrate. The VK-X150 was also used as a reflected light microscope and a Meiji MT9930L was used to perform dark-field microscopy. The thicknesses of the NCD films were measured with higher resolution than during growth using a Hamamatsu C13027 optical nano gauge. A FEI Quanta 250 FEG and a JEOL JSM-7900F SEM were used to examine the blind holes, the NCD films, and the TGVs. Grazing incidence X-ray diffraction measurements were carried out on the NCD films with a Bruker D8 Discover diffractometer, using Cu K$_{\upalpha}$ X-rays of 0.15418~nm average wavelength. The angle of incidence $\beta$ of the X-ray beam with the samples was 0.5$\degree$, which is slightly above the critical angle $\beta_c = 0.27 \degree$ of total external reflection for an diamond--air interface. Raman spectra were measured on a Tokyo Instruments Nanofinder 30 system. Specifically, a 5~mW 532~nm laser was focused by an objective lens on the sample to a spot of a few micrometers in diameter, which magnifies 100 times and has a numerical aperture of 0.95. Pressure was applied to the platform as described in previous work \cite{Janssens2014a}. Functions were fitted to data using the lmfit Python library.
\section{Results and discussions}

\subsection{Glass etching}
\begin{figure*}[t]
    \centering
    \includegraphics[scale = 0.95]{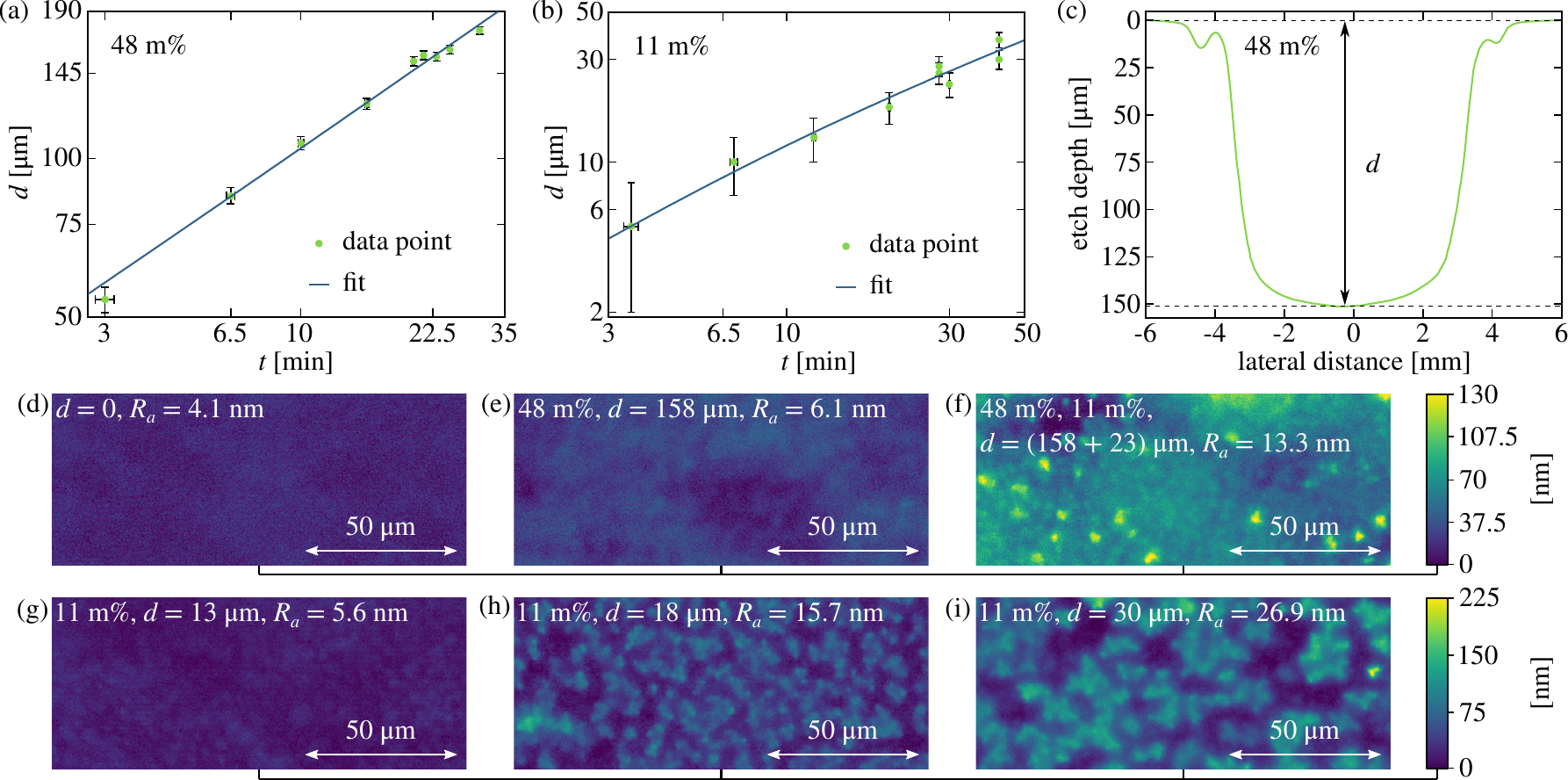}
    \caption{(a) Log--log graph of the maximum etch depth $d$ reached in a Lotus NXT glass substrate by HF etching versus etching time $t$. The substrate is etched locally around its center with 48~m\% HF. Each data point is obtained from one experiment and the least-squares method is used to fit Eq.~\ref{d_t_relation} to the data. The values of the fitting parameters $a$ and $b$ so obtained are $-4\pm 10~\upmu$m and $1054\pm79~\upmu$m$^2$~min$^{-1}$, respectively. (b) Data for the same experiment as in (a) but with 11~m\% HF. By fitting Eq.~\ref{d_t_relation} to the data, the values of fitting parameters $a$ and $b$ are $29\pm19~\upmu$m and $49\pm 17~\upmu$m$^2$~min$^{-1}$, respectively. (c) Surface profile of a substrate that is locally etched by HF with $d$ approximately 150~$\upmu$m. (d--f) Surface profiles of a glass substrate before etching, after etching to a depth of 158~$\upmu$m with 48~m\% HF during step~1, and after additionally etching 23~$\upmu$m  of the substrate with 11~m\% HF during step~4, respectively. The symbol $R_a$ denotes the arithmetic average roughness of the surface. (g--i) Surface profiles of glass substrates after etching 13~$\upmu$m, 18~$\upmu$m, and 30~$\upmu$m of pristine glass substrates with 11~m\% HF, respectively.}
    \label{data_etching}
\end{figure*}

We performed systematic studies to determine the etching times for steps 1 and 4. The results for 48~m\% and 11~m\% HF are presented in Figs.~\ref{data_etching}.a and b respectively, which show the maximum etch depth $d$ that is reached as a function of time $t$ for each of these concentrations. Each data point on the log--log graphs represents one experiment and $t$ and $d$ have respective errors of approximately $10$~s and 3~$\upmu$m. Fig.~\ref{data_etching}.c shows a typical surface profile taken after etching a substrate. To predict the time $t$ that is necessary to etch a certain depth $d$, we adapt a model that was first used to describe the oxidation of silicon \cite{Deal1965}. The model assumes steady-state diffusion, a first order chemical reaction, and a concentration of HF $c_0$ at $d = 0$ that is constant with $t$. Under these assumptions, the time-evolution of the depth is
\begin{equation}
d(t) = \frac{\sqrt{a^2 + 4 b t} - a}{2},
\label{d_t_relation}
\end{equation}
with
\begin{equation}
a = \frac{2D}{k} \qquad \text{and} \qquad b = \frac{2c_0D}{c_g},
\label{fitting_params}
\end{equation}
where $c_g$, $k$, and $D$ respectively denote the product of a stoichiometric constant and the concentration of species in glass that react with HF, the reaction constant, and a diffusion coefficient. Since most of these parameters are unknown, we fit the data in Figs.~\ref{data_etching}.a--b with Eq.~\ref{d_t_relation}. Results are provided in the caption of Fig.~\ref{data_etching}. We find that step~1 and step~4 can be performed in no more than 25~min and 35~min, respectively. During etching, a crust of fluorides forms on the surface of the glass substrate. After etching, that crust is removed by rinsing with deionized water.

Using Eq.~\ref{d_t_relation}, it can be found that $d \propto t$ for a reaction limited process and that $d \propto \sqrt{t}$ for a diffusion limited process \cite{Deal1965}. When fitting the general relation for a power law to the data for 48~m\% HF etching, we find that the power of $t$ is $0.48 \pm 0.2$, from which we infer that the etching process is diffusion limited. For 11~m\% HF etching, we find that the power of $t$ is $0.72 \pm 0.08$, which indicates that the etch process is on the border between reaction limited and diffusion limited. To understand the effect of HF concentration and duration on the etching process, we exploit the Damk\"{o}hler number
\begin{equation}
\textit{Da} = \frac{kd(t)}{D} = \sqrt{1 + 2 \frac{c_0 k^2 t}{c_g D}} - 1.
\label{Damkohler}
\end{equation}
This dimensionless quantity relates the reaction rate to the diffusion rate as follows: for $\textit{Da} \gg 1$, the etching process is diffusion limited; for $\textit{Da} \ll 1$, the etching process is reaction limited. From Eq.~\ref{Damkohler}, we find that the glass etching process is reaction limited for relatively small values of $t$ but becomes diffusion limited for relatively large values of $t$. We also infer that, consistent with our observations, diffusion limited behavior is achieved more rapidly for higher initial concentrations of HF.

The effect of HF etching on the surface roughness $R_a$ of a glass substrate was also investigated. This was done because rough surfaces scatter light, adversely affecting transparency. Prior to etching, the value of $R_a$ for a glass substrate was approximately 4~nm or less. During etching in step~1, $R_a$ increased up to 6~nm in the vicinity where $d$ was measured. Moreover, during etching in step~4, $R_a$ increased to approximately 13~nm. This cumulative increase in $R_a$ is evident in Figs.~\ref{data_etching}.d--f, which depict the surface profile of the glass substrate before step~1, between steps~1 and 2, and after step~4, respectively. It is remarkable that $R_a$ increased strongly after step~4, during which only approximately 25~$\upmu$m was etched. To provide additional proof that etching Lotus NXT glass with 11~m\% HF induced more surface roughness than with 48~m\% HF, we present in Figs.~\ref{data_etching}.g--i the surface profiles of glass substrates etched with 11~m\% HF to depths of approximately 13~$\upmu$m, 18~$\upmu$m, and 30~$\upmu$m. More surface profiles of etched glass substrates can be found as supplementary material. The mechanism underlying this observation is not yet understood. Assuming that masking of insoluble fluorides, which are sedimented at the HF--glass interface, causes roughening during etching, we hypothesize that for etching with 48~m\% HF, the velocity of the HF--glass interface is larger than the sedimentation rate of fluorides. Since step~4 required shallow etching with 11~m\% HF, the surface roughness of the glass substrate after etching was relatively low.

\subsection{Blind holes}
%
\begin{table}[b]
\footnotesize
  \caption{\ The depth of the blind holes that are depicted in Fig.~\ref{blind_holes}.a. See the caption of that figure for a detailed explanation of the procedure of making the holes. The error in measuring the depth is approximately 2~$\upmu$m. $P$ denotes the laser power scaled with the maximum power of the laser used in the experiment and $f$ denotes the depth measured from the etched surface of the substrate to the focal plane of the laser light in air.}
  \centering
  \label{ablation_table}
  \begin{tabular*}{0.48\textwidth}{@{\extracolsep{\fill}}l l l l l l}
     \toprule
     \multicolumn{6}{c}{\textbf{Depth of a blind hole [$\bm{\upmu}$m]}}\\
     \cmidrule{1-6}
     \multirow{2}{*}{$P$} & \multicolumn{5}{c}{$f$ [$\upmu$m]}\\
     & 15 & 20 & 25 & 30 & 35\\
     \cmidrule{2-6}
     0.36 & 21 & 24 & 28 & 31 & 33\\
     0.38 & 25 & 28 & 32 & 36 & 39\\
     0.40 & 28 & 31 & 35 & 38 & 48\\
     0.42 & 31 & 34 & 38 & 43 & 47\\
     0.44 & 34 & 38 & 42 & 47 & 49\\
     \bottomrule
  \end{tabular*}
\end{table}

\begin{figure*}
    \centering
    \includegraphics[scale = 0.95]{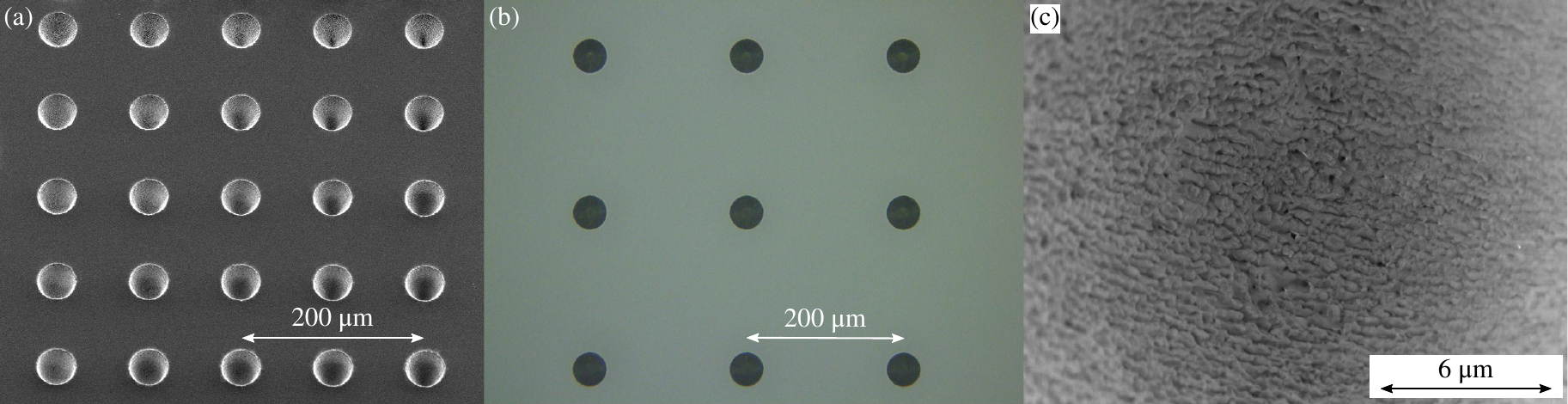}
    \caption{(a) SEM image at a tilt of 25$\degree$ depicting 25 blind holes of diameter 42~$\upmu$m in a Lotus NXT glass substrate. To make the holes, the surface of the substrate is placed at the focal pane $F$ of the laser light in air and a pattern of concentric circles is written 5 times. $F$ is then lowered to 5~$\upmu$m and the pattern is written 5 times again. This procedure is repeated till $F$ reaches a predefined depth $f$ measured from the etched surface of the substrate to $F$. From left to right, with respect to the figure, $f$ increases linearly from 15~$\upmu$m to 35~$\upmu$m and from top to bottom, and the laser power $P$, scaled with the maximum laser power of the system used in the experiment, increases linearly from 0.36 to 0.44. The depth of the blind holes is given in Table~\ref{ablation_table}. (b) Reflected light microscope image depicting 9 blind holes of diameter 42~$\upmu$m and depth 36~$\upmu$m that are made with $f=30$ and $P=0.38$, where the substrate is etched to a thickness of 46~$\upmu$m. (c) SEM image depicting the bottom of a blind hole. The hole is similar to those depicted in (b).}
    \label{blind_holes}
\end{figure*}

During the process of optimizing the parameters for laser ablation, we observed that crack formation becomes more apparent with increasing laser pulse frequency and $P$. To create blind holes at the lowest value of $P$ possible and thereby avoid crack formation, we performed a study in which the depth of a blind hole was measured as a function of $P$ for various choices of depth $f$. Results from this study are shown in Table~\ref{ablation_table} and an SEM image of the holes, taken under a tilt of 25$\degree$, from which the data is obtained, is shown in Fig.~\ref{blind_holes}.a. The procedure of making the holes is explained in the caption of that figure. From this study, it is clear that for $f=35$~$\upmu$m the lowest value of $P$ at which blind holes of approximately 40~$\upmu$m deep can be made is 0.38. Fig.~\ref{blind_holes}.b shows a reflected light microscope image of an array of 9 blind holes of approximately 42~$\upmu$m diameter and approximately 36~$\upmu$m deep in a substrate that is locally etched to a thickness of 46~$\upmu$m. These blind holes were made with $f=30$ and $P=0.38$. The values listed in Table~\ref{ablation_table} are visualized by a graph provided as supplementary material. Fig.~\ref{process}.d shows an SEM image, taken under a tilt of 25$\degree$, of a blind hole similar to those in Fig.~\ref{blind_holes}.b. From the image, it is clear that the shape of the blind hole resembles that of a paraboloid. This is most likely a consequence of laser light shading by the glass substrate. Fig.~\ref{blind_holes}.c shows an SEM image of the surface at the deepest point of the blind hole. Cracks caused by laser ablation are not detected from this image. After laser ablation, the thickness of the substrate at the deepest point of a blind hole was approximately 10 $\upmu$m. It is important to recognize that the the growth side does not show damage due to the laser ablation process.

\subsection{Nanocrystalline diamond}
\begin{figure*}
    \centering
    \includegraphics[scale = 0.95]{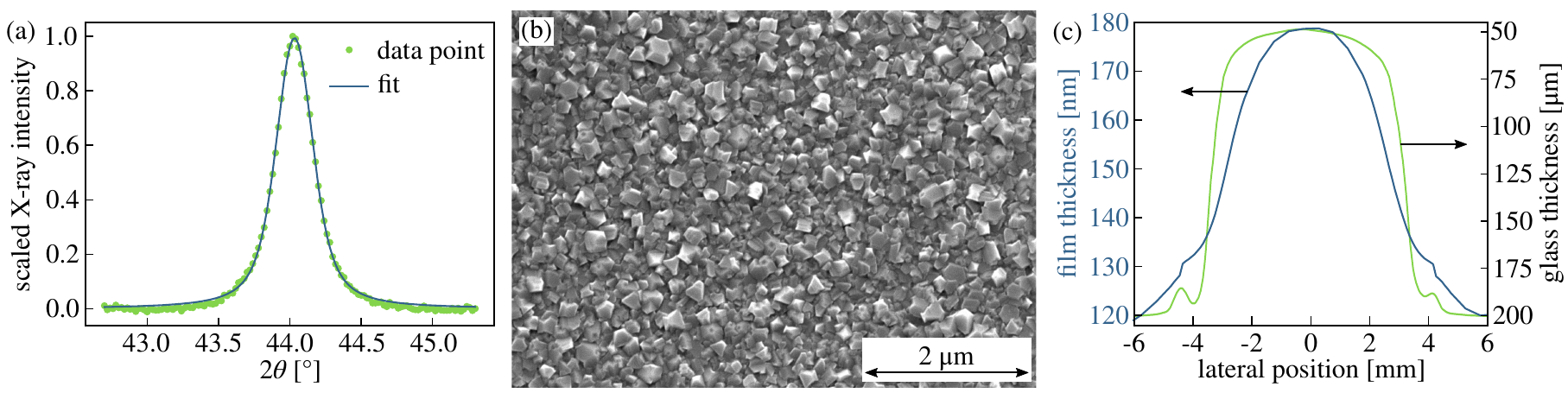}
    \caption{(a) Asymmetric grazing incidence X-ray diffractogram of the NCD film grown in step~3 with a Voigt function fitted to the data. The intensity of the scattered X-rays is scaled and $\theta$ denotes the Bragg angle. The center of the peak is at $2\theta = 44.03 \pm 0.01$$\degree$, which corresponds to constructive interference of copper K$_{\upalpha}$ X-rays that are scattered by the (111) crystal planes of diamond \cite{Bindzus2014}. (b) SEM image of the NCD film. The image shows that the film is pinhole-free and consists of nanoscopic diamond crystals. (c) Thickness of the NCD film and the thickness of the Lotus NXT glass substrate on which the film is grown. Both thicknesses are given as a function of a lateral position.}
    \label{diamond}
\end{figure*}
In step~3 of our process, an NCD film was grown on the surface of a glass substrate with blind holes on the growth side. During growth, the etching side was not in direct contact with the cooled substrate holder of the CVD system. An X-ray crystallography study, from which the results are shown in Fig.~\ref{diamond}.a, confirms that crystalline films were obtained. From the lattice parameters of diamond \cite{Bindzus2014}, we deduce that the peak is caused by (111) crystal planes of diamond that are compressively strained by 0.1\%. Results from a Raman spectroscopy study that is included in the supplementary material support these findings. A diffractogram for the (220) and (311) peaks, which are about an order of magnitude smaller than the (111) peak, is not provided due to long integration times. Fig.~\ref{process}.e shows a photograph of a substrate after NCD growth that was taken with the camera located on the growth side. The substrate was placed on a mirror and therefore a large part of the light recorded by the camera was scattered, making it possible to observe the array of 9 blind holes. Fig.~\ref{diamond}.b shows an SEM image of the NCD film, which exhibits a pinhole-free nanocrystalline structure. SEM images of this film taken with different magnifications are provided as supplementary material. Fig.~\ref{diamond}.c show the thickness of the film together with the thickness of the glass substrate. At the thinnest part of the glass substrate, in the vicinity of the blind holes, the film shown is $175\pm5$~nm and reduces from that area radially towards the edges of the diamond film to approximately $120\pm5$~nm. Due to thin-film interference \cite{Born2002}, the variations in film thickness are observable from differences in the apparent color of the NCD film, as shown in Fig.~\ref{process}.e.

To explain variations in the thickness of the NCD film, which is depicted in Fig.~\ref{diamond}.c, we note that, during NCD growth, only the glass substrate surface on the etching side, where no thinning is performed, is in contact with the water-cooled substrate holder whilst the glass substrate surface on the opposite growth side is in contact with a plasma. It is therefore reasonable to assume that during growth, the temperature $T_s$ of the etched part of the glass substrate is greater than that of the remainder of the substrate. Following Tsugawa et al.\ \cite{Tsugawa2010}, who claim that the growth rate $r$ of diamond increases monotonically with $T_s$, following an Arrhenius-type equation
\begin{equation}
r = A e^{-E_a/k_BT_s},
\label{growth_rate}
\end{equation}
where $A$, $E_a$, and $k_B$ respectively denote a pre-exponential factor, the activation energy, and the Boltzmann constant, we propose that the growth rate of the NCD film is spatially nonuniform. Since the growth time at all locations is the same, a nonuniform film thickness is expected. By growing the NCD film before etching, variations in film thickness can be avoided. However, it is known that NCD can be transformed into CO$_2$ at temperatures above $550 \pm 50 \degree$C \cite{Kriele2011}. The question of whether or not the heat generated during the laser ablation process induces such temperatures and affects the NCD remains to be investigated. It is worth noting that the substrate is an oxide and therefore barely interacts with air. Also, the softening point of the substrate lays at $1043\degree$C, which makes it resilient against relatively high temperatures.

\subsection{Through glass vias}
\begin{figure*}
    \centering
    \includegraphics[scale = 0.95]{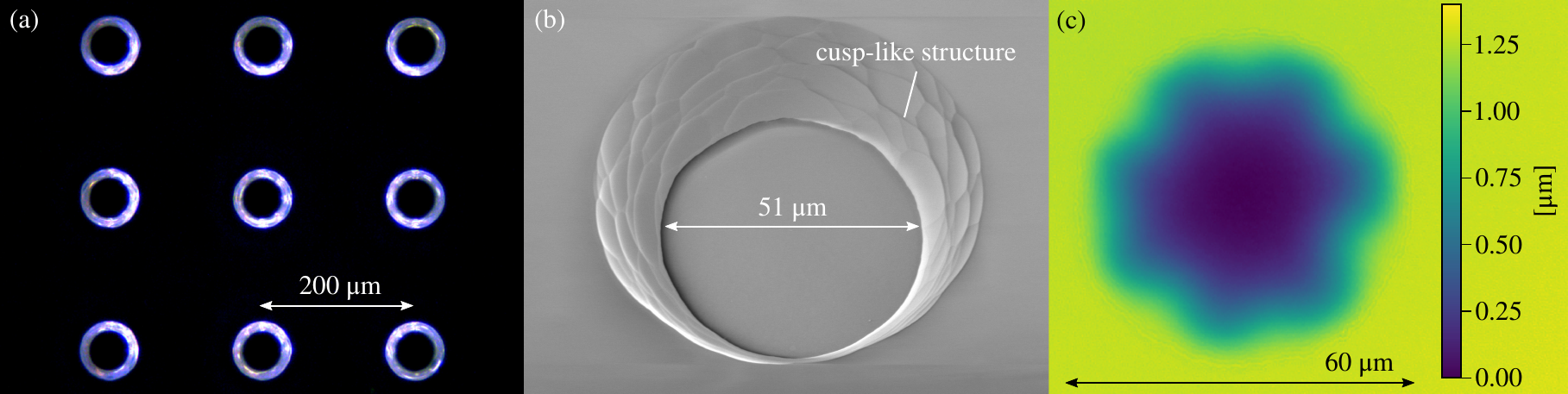}
    \caption{(a) Dark-field optical micrograph of the Lotus NXT glass substrate with NCD sealed TGVs that are created by our process. The image captures the same NCD-glass platform as that appearing in Fig.~\ref{process}.f. The image shows that apart from the walls of the TGVs, the platform barely scatters light. (b) SEM image of the center TGV, which is taken on the etching side under a tilt of 25$\degree$. The wall of the TGV is clearly rougher than the rest of the NCD-glass platform. (c) Surface profile of a portion of the NCD film that seals the center TGV, taken on the growth side of the glass substrate.}
    \label{tgv}
\end{figure*}

In step~4 of our process, the glass substrate is further etched by HF to form through glass vias (TGVs) that are sealed with suspended portions of an NCD film of approximate diameter 60~$\upmu$m. Fig.~\ref{process}.f shows a micrograph, which is taken on the etching side, of nine TGVs sealed with suspended NCD of thickness $175\pm5$~nm. In the vicinity of these TGVs, the substrate is of approximate thickness 23~$\upmu$m and light scattering causes the walls of the TGVs to appear as black circles. Fig.~\ref{tgv}.a is a dark-field optical microscope image, taken on the growth side of the substrate, that captures the same NCD-glass platform as Fig.~\ref{process}.f. From this image, we infer that, despite the walls of the TGVs, the platform barely scatters light. Fig.~\ref{process}.g shows a micrograph taken on the growth side. The circular shape represents the suspended portion of the NCD film. Fig.~\ref{tgv}.b depicts an SEM image of the center TGV, which is taken on the etching side of the substrate, under a tilt of 25$\degree$. The wall of the TGV is clearly rougher than the rest of the structure. All TGVs are tapered from the etching side towards the growth side and have minimum and maximum diameters of approximately 50~$\upmu$m and 80~$\upmu$m, respectively. Since the diameter of the suspended film is approximately 60~$\upmu$m, the NCD film is underetched by approximately 5~$\upmu$m. Fig.~\ref{tgv}.c depicts the surface profile of a portion of the NCD film that seals the central TGV. The profile is taken on the growth side and the suspended portion of the NCD film is slightly buckled towards the glass substrate with a maximum deflection of approximately 1.25~$\upmu$m. This behavior is seen for all suspended portions of the NCD film that seal TGVs, as can be seen in the supplementary material. To show that the bond between the NCD film and the glass substrate is sufficiently strong for practical use, we applied a gauge pressure of 300~kPa to the suspended NCD on the etching side, the result of which was that the structure of the NCD-glass platform was able to withstand such pressures.

After laser ablation in step~2 of our process, we did not observe cracks in the blind holes; however, cusp-like structures developed on the walls of a TGV, as shown in Fig.~\ref{tgv}.b.  Spierings~\cite{Spierings1993} showed that such structures typically develop from cracks during glass etching. This suggests that any cracks during the laser ablation process were too small to be detected by our imaging techniques. Still, due to the optimization of the laser ablation process, we achieved a feature size that allows the TGVs to be fairly circular. It strikes us as highly likely that using an excimer laser for laser ablation  \cite{Sukumaran2012,Sato2019} or activating the glass with a laser instead of performing laser ablation \cite{Jaramillo2018} would further reduce the roughness of the TGVs after etching. More details on modifying glass with laser light can be found in the literature \cite{Sugioka2014}.

During preliminary experiments, we found that NCD films, grown on fused silica substrates during step~3 of our process, completely delaminated during step~4. We suspect that this is caused by tensile stress, which is to be expected since the coefficient of thermal expansion of fused silica is less than that of diamond over the entire range of operating temperatures used during our preliminary experiments \cite{Moulle1997,Oishi1969}. Our experiments and previous work \cite{Janssens2014a,Janssens2014b,Drijkoningen2016} show that NCD films grown on substrates made of any glass listed in Table~\ref{types_of_glass}, except fused silica, are compressively stressed, which leads us to suspect that the coefficients of thermal expansion of those types of glass might be greater than that of diamond for significant portions of the temperature ranges used in the relevant growth processes. It is noteworthy that Iliescu et al.~\cite{Iliescu2012} and Ceyssens and Puers~\cite{Ceyssens2009} found that compressive stress acting on films, which mask parts of glass substrates during HF etching, is preferable over tensile stress. Our findings agree with their results. A typical consequence of compressive stress is buckling \cite{Thouless1991}, which is illustrated by Fig.~\ref{tgv}.c. For several applications, buckling should be minimized by properly tuning the coefficients of thermal expansion of the film and the substrate. Since the coefficient of thermal expansion of diamond exceeds that of fused silica but is less than those of many other glasses, we are optimistic that this can be done.

\section{Conclusion}
We present a low-cost and robust nanocrystalline diamond-glass platform for single-cell culture and analysis, on-demand drug delivery systems, the modeling of vascular systems, microelectrodes, quantum technologies, and high-temperature MEMS. Our platform is comprised of a glass substrate with through glass vias (TGVs) that are sealed on one side with suspended portions of an ultra-thin nanocrystalline diamond (NCD) film. Our fabrication process is free of photolithography and transfer printing and is delineated in detail sufficient to allow easy replication by others. In this process, hydrofluoric acid (HF) is used to first etch one side of a $10\times10\times0.2$~mm$^3$ Lotus NXT glass substrate to a thickness of approximately 50~$\upmu$m. On the same side that is etched, blind holes of approximate diameter $40$~$\upmu$m and approximate depth $40$~$\upmu$m are subsequently formed by laser ablation. After growing an NCD film of approximate thickness 175~nm on the surface opposite to the etched side, the etched side of the substrate is further etched by HF to approximately 25~$\upmu$m to produce the NCD sealed TGVs. Our resulting platform is highly transparent and can handle applied pressures of at least 300~kPa.

\section*{Acknowledgements}
We gratefully acknowledge support from the Okinawa Institute of Science and Technology Graduate University with subsidy funding from the Cabinet Office, Government of Japan.

\bibliography{library}
\end{document}


\maketitle
\vspace{-1cm}
\begin{center}
Mathematics, Mechanics, and Materials Unit (MMMU), Okinawa Institute of Science and Technology Graduate University (OIST), 1919-1 Tancha, Onna-son, Kunigami-gun, Okinawa, Japan 904-0495
\end{center}
{\let\thefootnote\relax\footnote{Email address: stoffel.janssens@oist.jp (S.\ D.\ Janssens).}}
\setcounter{footnote}{0}
\vskip15pt
\section{Chemical reactor}
\begin{figure*}[h]
    \centering
    \includegraphics{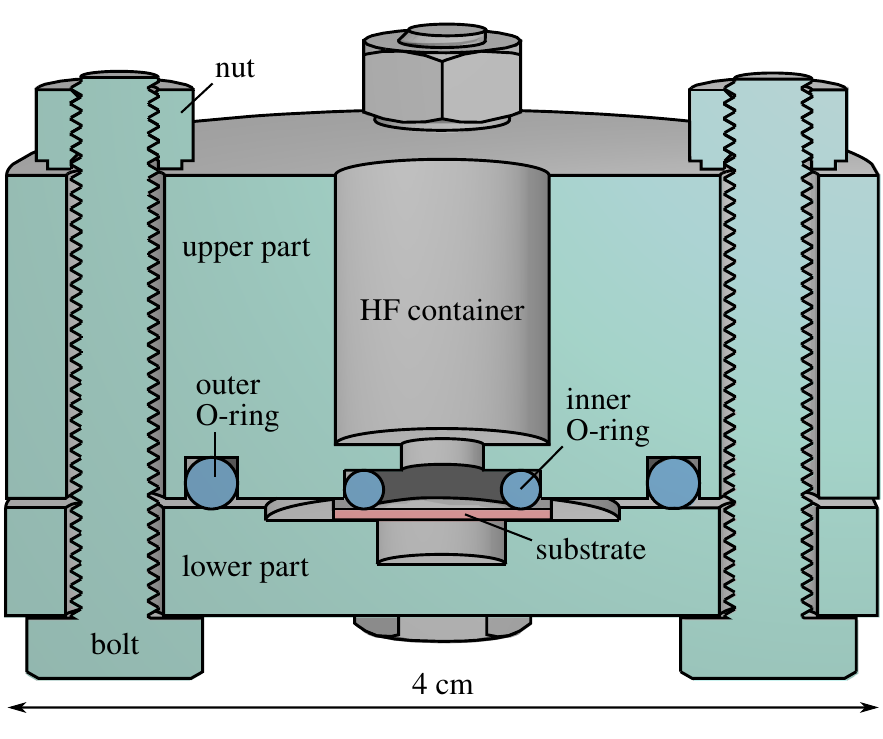}
    \caption{Cross-section of a schematic of the home-made chemical reactor used for etching a glass substrate by hydrofluoric acid.}
    \label{chemical_reactor}
\end{figure*}

Supplementary Figure~\ref{chemical_reactor} shows a cross-section of the schematic of the home-built chemical reactor used for etching a glass substrate by hydrofluoric acid (HF). The cross-section is taken at a mirror plane of the reactor. The reactor is designed to partly expose one surface of a substrate to HF. By tightening the four nuts and bolts, the inner O-ring of the reactor presses the upper part of the reactor against the surface of the substrate that is in contact with the lower part of the reactor. In the upper part of the reactor, a through hole is present that acts as a container in which up to 0.8~ml of the HF can be placed. The inner O-ring sits around the through hole, has an inner diameter of 5.8~mm, and keeps the HF in the container. The outer O-ring is clamped by the four screws between the upper part and the lower part of the chemical reactor, has an inner diameter of 17.8~mm, and acts as a safety barrier against leakage of the HF. The O-rings are made of perfluoro rubber and all other parts of the reactor are made of polytetrafluoroethylene (PTFE). Both perfluoro rubber and PTFE are strongly resistant against HF.
\newpage

\section{Glass substrates: profilometry}
\begin{figure*}[h]
    \centering
    \includegraphics[scale = 0.95]{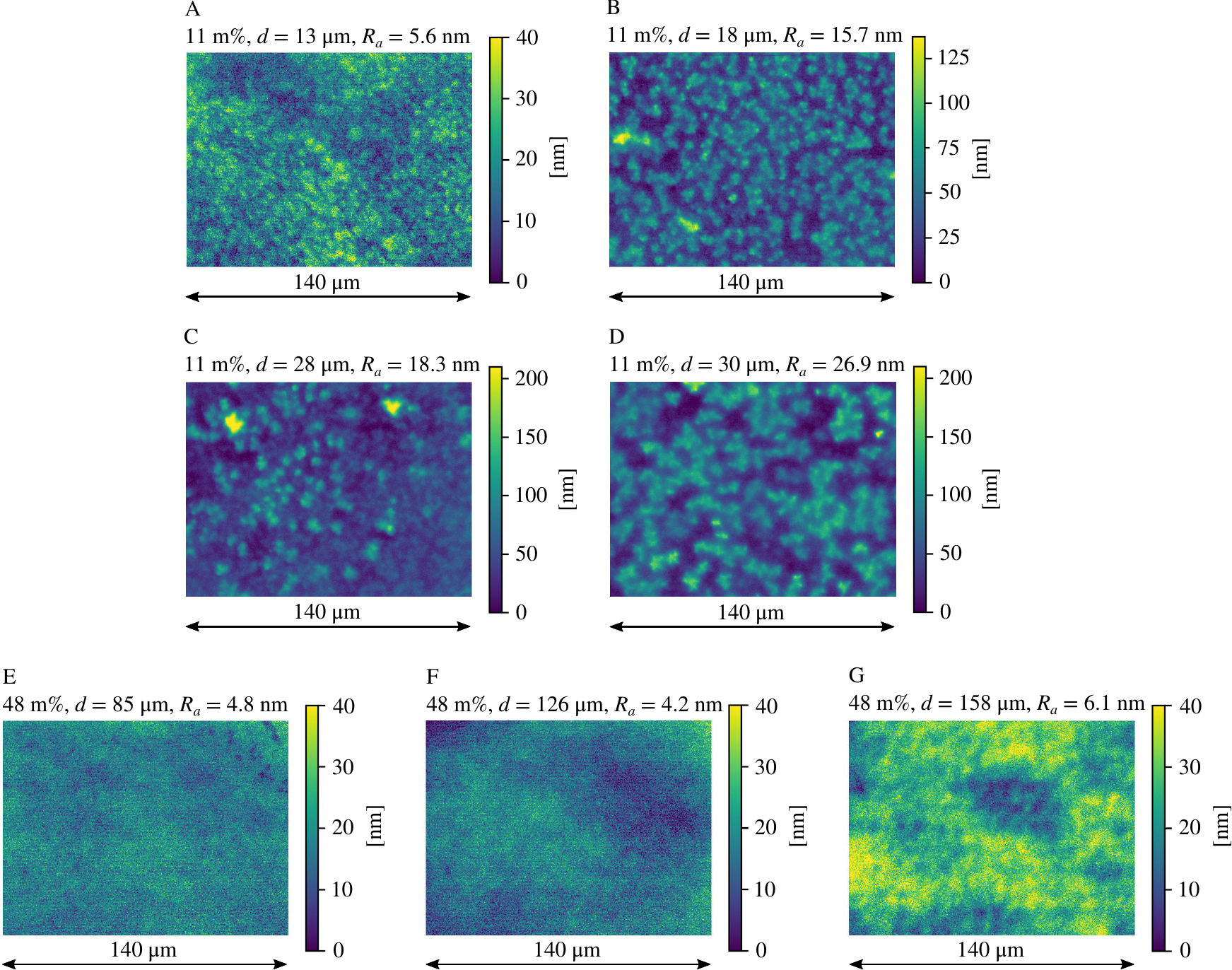}
    \caption{A--D) Surface profiles of Lotus NXT glass substrates etched with 11~m\% HF. E--G) Surface profiles of Lotus NXT glass substrates etched with 48~m\% HF. Above each profile, the etch depth $d$ and the surface roughness $R_a$ are given.}
    \label{roughness}
\end{figure*}

Supplementary Figures~\ref{roughness}A--D depict surface profiles of Lotus NXT glass substrates that are etched with 11~m\% HF. Supplementary Figures~\ref{roughness}E--G depict surface profiles of Lotus NXT glass substrates that are etched with 48~m\% HF. The profiles are formed after reaching various etch depths $d$ as indicated and, for each profile, the surface roughness $R_a$ is given. We conclude that 11~m\% HF roughens the surface of Lotus NXT glass substrates more than 48~m\% HF does.
\newpage

\section{Blind holes}
\begin{figure*}[h]
    \centering
    \includegraphics[scale = 0.95]{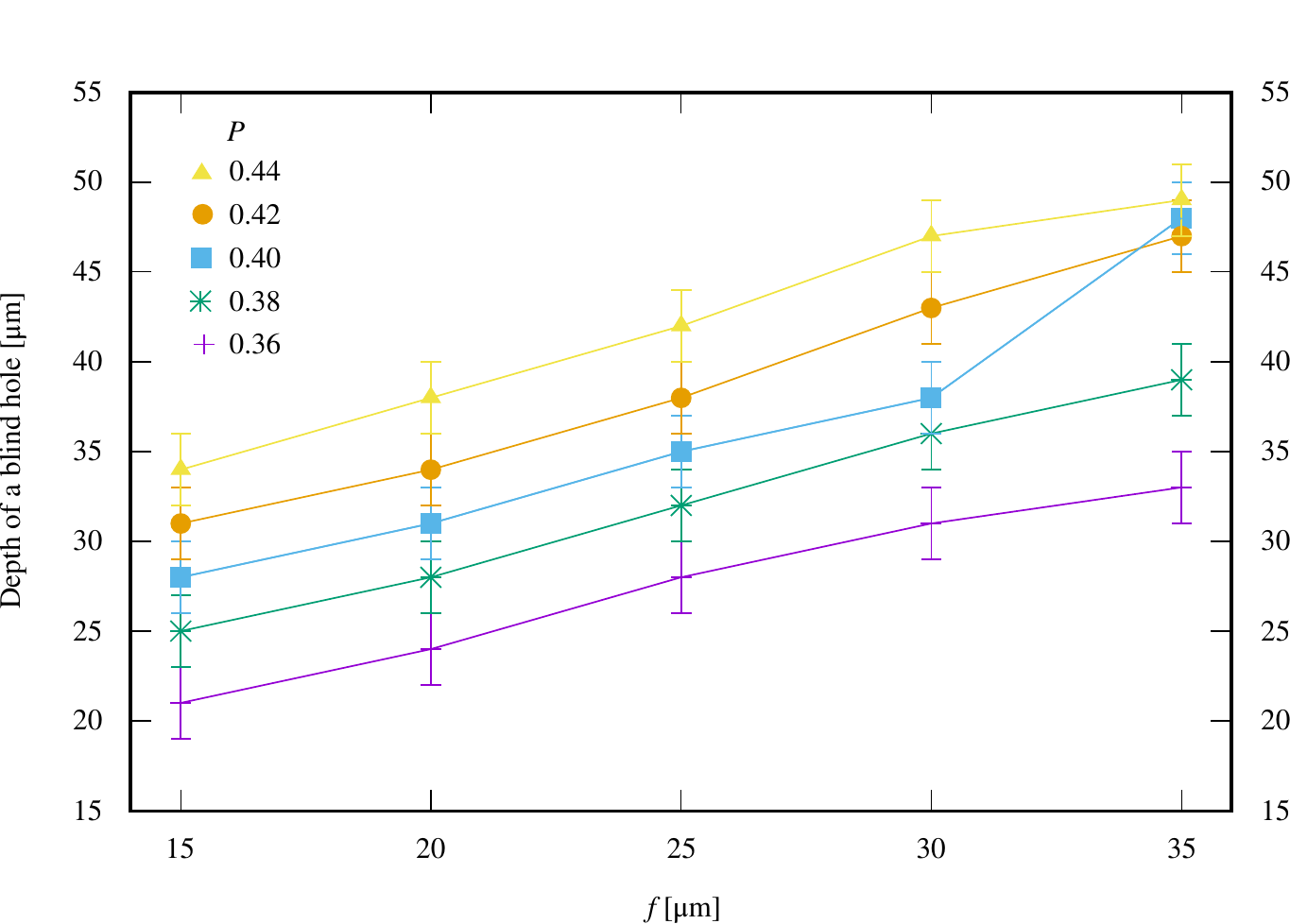}
    \caption{The depth of 25 blind holes of 42~$\upmu$m diameter that are created by laser ablation in a $10 \times 10 \times 0.2$~mm$^3$ Corning Lotus NXT glass substrate on the etching side of the substrate. The depth of a blind hole is given as a function of the depth $f$ measured relative to the etched surface of the substrate and the laser power $P$ scaled with the maximum power of the laser.}
    \label{laser_ablation}
\end{figure*}

Supplementary Figure~\ref{laser_ablation} shows the depth of 25 blind holes of 42~$\upmu$m diameter that are made using the front side laser ablation technique in a $10 \times 10 \times 0.2$~mm$^3$ Corning Lotus NXT glass substrate on the etching side of the substrate, with the etching side denoting the side where the substrate is etched in steps~1 and 4 of our process. The holes are made in the vicinity where the glass substrate is etched most and is of approximate thickness 50~$\upmu$m. To create these holes, the focal plane $F$ of the laser light in air is placed at the surface of the substrate and a pattern of concentric circles is written 5 times. $F$ is then lowered by 5~$\upmu$m, and the pattern is again written 5 times. This procedure is repeated till $F$ reaches a predefined target depth $f$, with $f$ denoting the vertical distance from $F$ to the etched surface. The symbol $P$ denotes the laser power scaled with the maximum power of the laser. The depth of each blind hole is taken from Table~2. It is clear that the depth of the blind hole that is made with $f=35$ and $P=40$ deviates from the expected value. So far, the cause of this is unclear. However, we expect that using the stage of the LightFab for changing $f$ rather than the galvo might be the source of this deviation.
\newpage

\section{Nanocrystalline diamond: scanning electron microscopy}
\begin{figure*}[h]
    \centering
    \includegraphics[scale = 0.95]{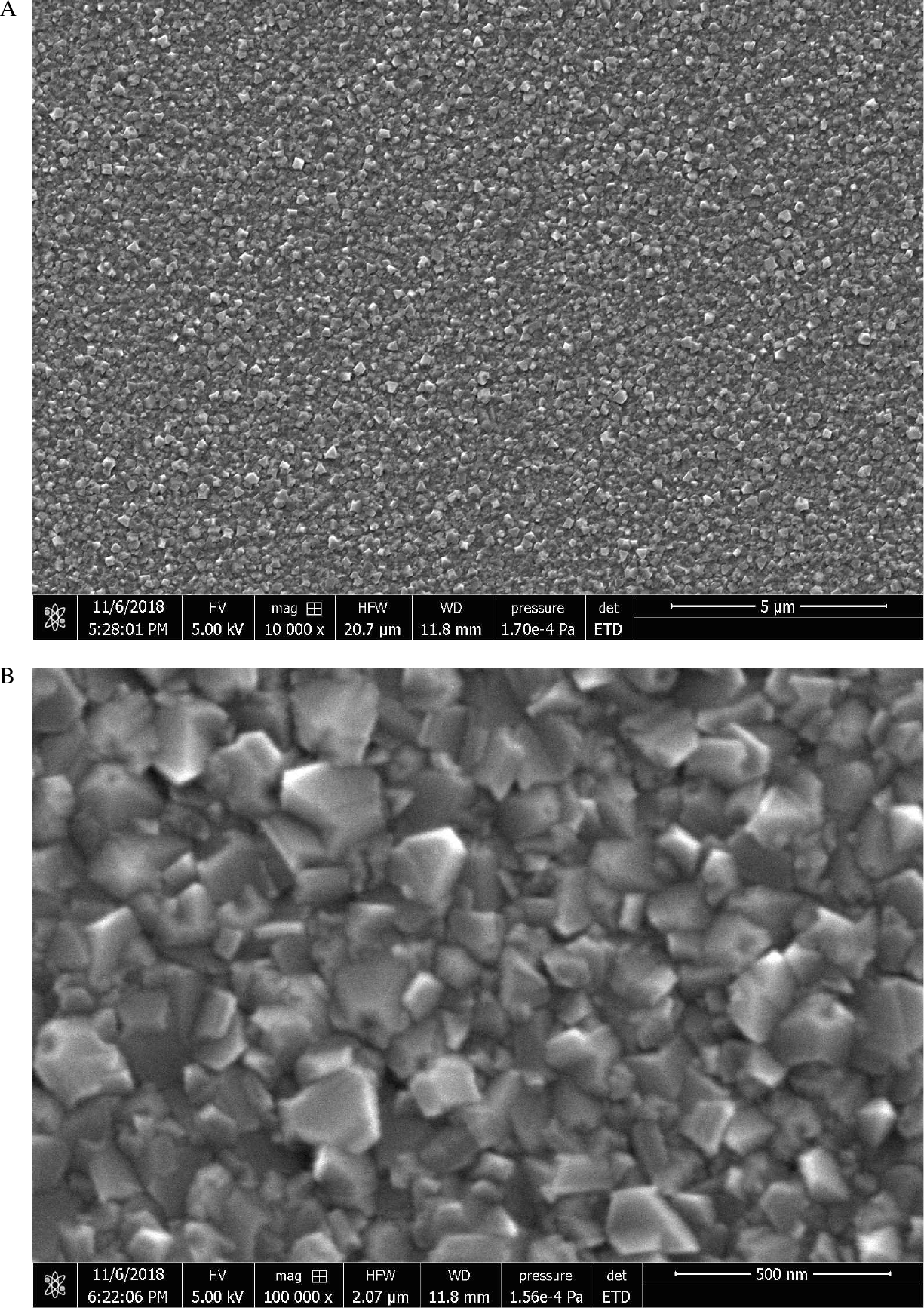}
    \caption{A) SEM image of an NCD film that is grown on a Lotus NXT glass substrate. The film is of approximate thickness 175~nm and magnified $10^4$ times. B) SEM image of the same NCD film but magnified $10^5$ times.}
    \label{SEM}
\end{figure*}

Supplementary Figures~\ref{SEM}A--B show scanning electron microscope (SEM) images at different magnifications of the nanocrystalline diamond (NCD) film that is depicted in Fig.~4.b. The image is taken at a location where the film is of approximate thickness 175~nm. The film is grown with microwave plasma-enhanced chemical vapor deposition in the reactor of an SDS6500X system on a Lotus NXT glass substrate after seeding the substrate with nanodiamonds.
\newpage

\section{Nanocrystalline diamond: Raman spectroscopy}
\begin{figure*}[h]
    \centering
    \includegraphics[scale = 0.95]{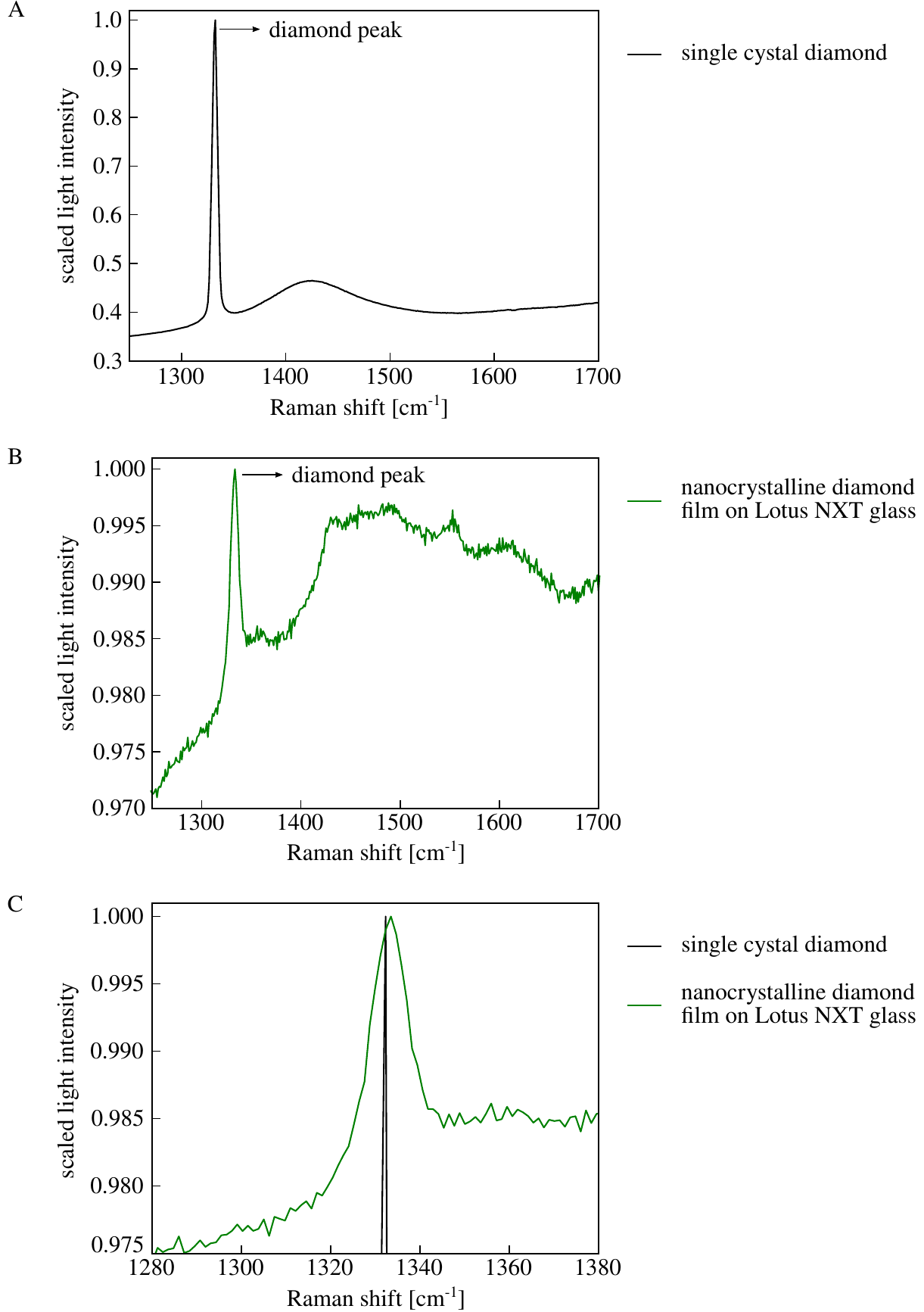}
    \caption{A) Scaled Raman spectrum of a single crystal diamond. B) Scaled Raman spectrum of the NCD film depicted in Fig.~4.b. C) Zoom of the diamond peaks that appear in A and B.}
    \label{Raman}
\end{figure*}

Supplementary Figures~\ref{Raman}A--B depict scaled Raman spectra of a (111)-oriented $3\times3\times0.3$~mm$^3$ single crystal diamond and of the NCD film depicted in Fig.~4.b that is grown on a Lotus NXT glass substrate. In each of these spectra, the most intense feature forms a peak that originates from the first-order Raman line of diamond, which is expected at the approximate Raman shift 1332~cm$^{-1}$.$^{1}$ Supplementary Figure~\ref{Raman}C shows a zoom of those peaks. It can be observed that the center of the diamond peak extracted from the NCD film is positioned at a shift of approximate value 1334~cm$^{-1}$, indicating that the film is strained due to a compressive stress of approximately 0.5~GPa acting on the film.\footnote{H. Windischmann, and K. J. Gray, \textit{Diamond Relat.\ Mater.}, 4:837--842, 1995.} The compressive stress is mainly caused by a thermal mismatch between NCD and the glass since a portion of unstrained freestanding film generates a diamond peak approximately at 1332~cm$^{-1}$. The large band located at shifts larger than 1332~cm$^{-1}$ and the large photoluminescence background observed in Supplementary Figure~\ref{Raman}B are mainly ascribed to sp$^2$ carbon and hydrogen that are present in the NCD film, respectively.\footnote{F.\ Klauser, D.\ Steinm\"uller-Nethl, R.\ Kaindl, E.\ Bertel, and N.\ Memmel, \textit{Chem.\ Vap.\ Deposition}, 16:127--135, 2010.} The presence of impurities as well as point and line defects are also a known factor for the broadening of the diamond peak, which explains the considerably wider peak obtained for NCD compared to the peak obtained for single crystal diamond.\footnote{L.\ Bergman, and R.\ J.\ Nemanich, \textit{J.\ Appl.\ Phys.} 78:6709--6719, 1995.} Finally, it is noteworthy that this analysis is consistent with the X-ray diffraction study reported in the main text.

\newpage

\section{Suspended nanocrystalline diamond: optical microscopy}
\begin{figure*}[h]
    \centering
    \includegraphics[scale = 0.95]{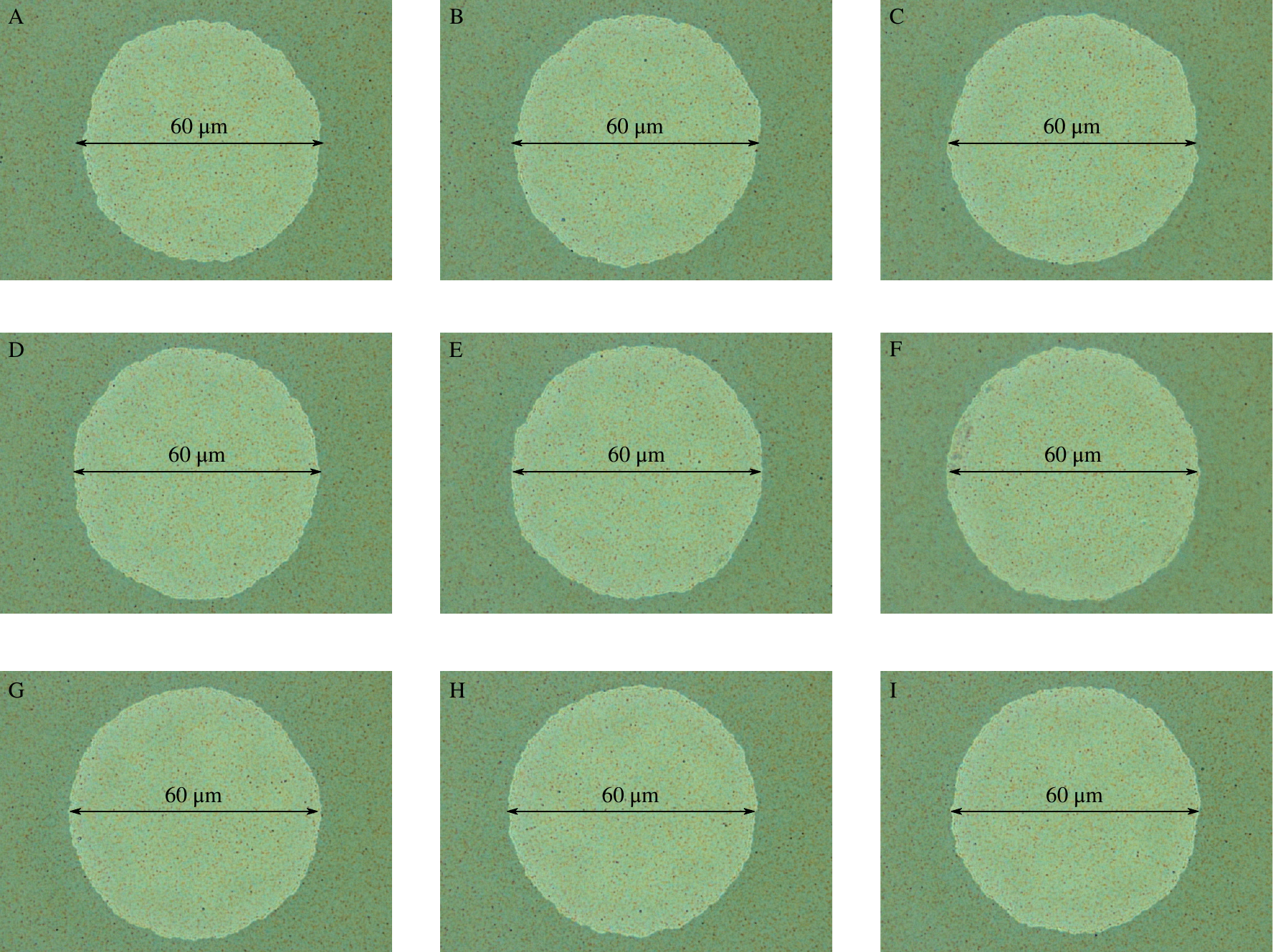}
    \caption{A--I) Reflecting microscope images, taken on the growth side of the glass substrate, of portions of the NCD film that seal the TGVs depicted in Fig.~1.f. The circular shapes represent the suspended portions of our structures.}
    \label{Sus_NCD}
\end{figure*}

Supplementary Figures~\ref{Sus_NCD}A--I show images of portions of an NCD film that seals the TGVs depicted in Fig.~1.f. These images are taken with a reflecting light microscope with the objective on the growth side of the glass substrate on which the NCD film is grown. A circular shape represents a suspended portion of the 175~nm thick NCD film. The shapes are observable due to a difference between the refractive index of air and glass. Supplementary Figures~\ref{Sus_NCD}A--I illustrate that structures of similar shape are fabricated with our process.
\newpage

\section{Suspended nanocrystalline diamond: profilometry}
Supplementary Figures~\ref{Surf_prof}A--I show the surface profiles of the portions of the NCD film that are depicted in Supplementary Figures~\ref{Sus_NCD}A--I. These profiles are taken on the growth side of the glass substrate and all suspended portions of the NCD film are slightly buckled towards the glass substrate with maximum deflections of approximately 1.25~$\upmu$m.
\begin{figure*}[h]
    \centering
    \includegraphics[scale = 0.95]{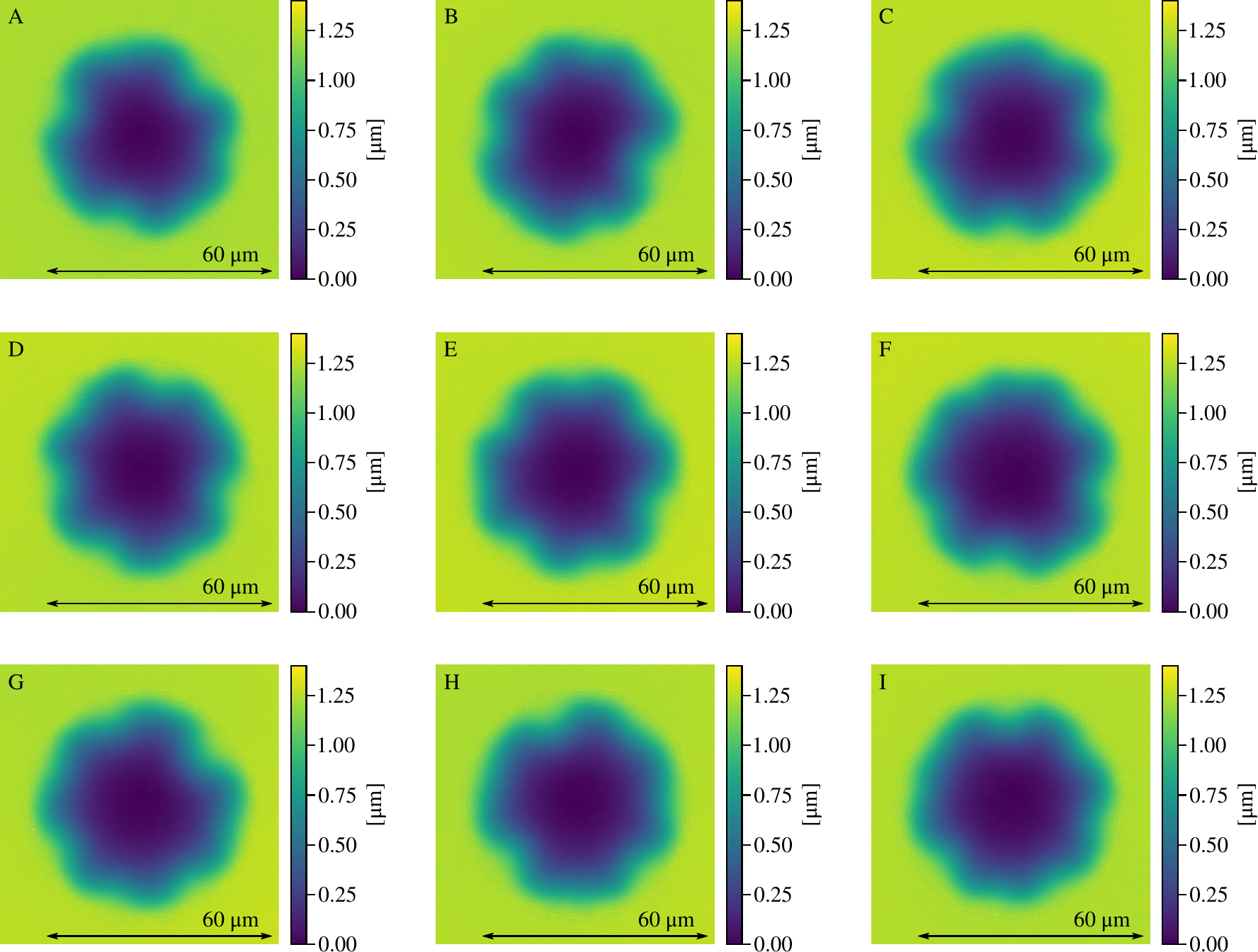}
    \caption{A--I) Surface profiles, taken on the growth side of the glass substrate, of the portions of the NCD film that are depicted in Supplementary Figures~\ref{Sus_NCD}A--I.}
    \label{Surf_prof}
\end{figure*}